\documentclass[aps,prd,twocolumn,showpacs,preprintnumbers,superscriptaddress,groupedaddress]{revtex4-1} 
\usepackage{dcolumn} 
\usepackage{graphicx}
\usepackage{slashed,color}
\usepackage{amsmath,amsfonts,bm}
\usepackage{amssymb}
\usepackage{calligra}
\usepackage[T1]{fontenc}

\newcommand{\nn}{\nonumber \\ }

\begin{document}


\title{Quasi-PDFs, momentum  distributions   and pseudo-PDFs}

\author{A. V. Radyushkin}

\affiliation{Old Dominion University, Norfolk,
             VA 23529, USA}
\affiliation{Thomas Jefferson National Accelerator Facility,
              Newport News, VA 23606, USA
}

\begin{abstract}

We  show that  quasi-PDFs may be  treated  as hybrids of PDFs and 
primordial rest-frame 
momentum distributions of partons. 
 This results in a   complicated  convolution  nature of quasi-PDFs that 
necessitates  using  large $p_3 \gtrsim 3$ GeV 
momenta  to 
 get reasonably close to the PDF limit.
 As  an alternative approach,
 we propose   to use 
pseudo-PDFs  ${\cal P}(x, z_3^2)$   that generalize
the light-front PDFs   onto spacelike intervals and  are related to 
 Ioffe-time  distributions  ${\cal M} (\nu, z_3^2)$, the  functions of   
  the Ioffe time  $\nu = p_3 z_3$ and the distance parameter $z_3^2$
  with respect to which it displays perturbative evolution for small $z_3$. 
 In this form, one may  
 divide   out  the   $z_3^2$   dependence  coming from the primordial rest-frame 
 distribution 
 and from the  problematic   factor  due to   lattice renormalization  of the  gauge link.  
 The $\nu$-dependence remains intact and determines the shape of PDFs.


              
\end{abstract}


\pacs{12.38.-t, 
      11.15.Ha,  
      12.38.Gc  
}

\maketitle


\section{ Introduction}

The parton distribution functions (PDFs) $f(x)$ \cite{Feynman:1973xc}
are  related to   matrix elements
of bilocal operators on   the light cone $z^2=0$, which prevents 
a straightforward calculation of these functions in  the   lattice 
 gauge  theory 
 formulated in  Euclidean space. 
  The usual way  out is to calculate  their moments. 
 However, recently, X. Ji  \cite{Ji:2013dva} suggested a method 
 allowing to calculate PDFs   as  functions of $x$. 
 To this end, he proposes to use purely space-like
separations $z=(0,0,0,z_3)$.  
Then one deals with  quasi-PDFs  $Q(y,p_3)$ describing sharing 
of   the $p_3$  hadron  momentum  component,  
and 
   tending  to  PDFs $f(y)$ 
     in the \mbox{$p_3 \to \infty$}  limit.
     The same method can be applied to distribution amplitudes (DAs).  
The results of lattice calculations of  
quasi-PDFs   were reported  
in Refs.  \cite{Lin:2014zya,Chen:2016utp,Alexandrou:2015rja}
and of the pion quasi-DA in Ref. \cite{Zhang:2017bzy}.

In our recent papers \cite{Radyushkin:2016hsy,Radyushkin:2017gjd}, 
we  have studied  nonperturbative $p_3$-evolution of
quasi-PDFs and quasi-DAs 
using the formalism of  virtuality distribution functions 
 \cite{Radyushkin:2014vla,Radyushkin:2015gpa}.
 We found that  quasi-PDFs can be obtained from 
  the   transverse momentum dependent 
 distributions    (TMDs) ${\cal F} (x, k_\perp^2)$. We  
 built models for the    nonperturbative evolution of
quasi-PDFs using simple models for TMDs. 
Our results are in qualitative agreement with the
\mbox{$p_3$-evolution}   patterns  obtained in lattice calculations.

In the present paper,  our  first goal 
 is to  develop   a picture for 
 quasi-PDFs as hybrids of PDFs and 
primordial 
momentum distributions of partons in a hadron at rest. 
As an intermediate step, we demonstrate  that the connection between 
TMDs and  quasi-PDFs
\cite{Radyushkin:2016hsy} is a mere consequence 
of  Lorentz invariance. Then we  show that, 
when a  hadron is moving, the parton $k_3$ momentum
comes from two sources. The motion
of the hadron as a whole gives 
the  $xp_3$  part,  governed by the dependence of
 the TMD  ${\cal F} (x, \kappa^2)$ on its $x$ argument.
  The remaining part $k_3-xp_3$   
is governed by the  dependence of the TMD  on its second argument,
$\kappa^2$, 
dictating    the primordial 
 rest-frame 
 momentum distribution.    The convolution  nature of quasi-PDFs 
 results in a rather complicated pattern of their 
 \mbox{$p_3$-evolution,} necessitating rather   large values $p_3 \sim 3$ GeV 
 for getting close to the PDF limit.
 
Thus, our second goal is to propose an alternative approach for 
lattice PDF extraction.   To this end, we introduce 
{\it pseudo-PDFs} $ {\cal P} (x, z_3^2) $  that generalize 
the light-cone PDFs $f(x)$ onto spacelike intervals like  
 $z=(0,0,0,z_3)$.   The pseudo-PDFs are Fourier transforms 
 of the  \mbox{{\it Ioffe-time}  \cite{Ioffe:1969kf}}  {\it distributions} \cite{Braun:1994jq}
 ${\cal M} (\nu, z_3^2)$ 
 that are  basically given by  generic matrix elements  like $  \langle p |   \phi(0) \phi (z)|p \rangle $
 written as  functions of $\nu = p_3 z_3$ and $z_3^2$. 
 Unlike quasi-PDFs, the pseudo-PDFs have the ``canonical''  $-1 \leq x  \leq 1$ support
 for all $z_3^2$.  They  tend to  PDFs when  $z_3\to 0$, 
  showing in this limit  a usual perturbative evolution 
 with $1/z_3$ serving as an evolution parameter. 
 Finally, we discuss  how  these properties of  pseudo-PDFs  may be used 
 for extraction of  PDFs on the lattice.

 \section{Parton distributions} 


  \subsection{Generic matrix element  and 
Lorentz invariance} 

  Historically  \cite{Feynman:1973xc},  PDFs were  introduced
  to describe  \mbox{spin-1/2 quarks. }
Since  complications related to spin
 do not affect  
  the very concept of parton distributions,  we  start with 
  a simple example of a  scalar  theory.
In that case,  information about the target is accumulated 
in the generic matrix element $\langle  p | \phi (0)  \phi(z)  | p \rangle $.
By Lorentz invariance, it is a function of two scalars, $(pz) \equiv - \nu$ and $z^2$
(or $-z^2$ if we want a positive value  for spacelike $z$):
  \begin{align}
  \langle p |   \phi(0) \phi (z)|p \rangle 
=  & {\cal M} (-(pz), -z^2) 
\,  .
 \
 \label{lorentz}
\end{align} 
 It can be shown \cite{Radyushkin:2016hsy,Radyushkin:1983wh}    that, for all contributing  Feynman diagrams,   
 its  Fourier transform  ${\cal P} (x, -z^2)$ with respect to $(pz)$ 
 has the $-1 \leq x \leq 1$ support, i.e., 
   \begin{align}
 {\cal M} (-(pz), -z^2) 
&   = 
 \int_{-1}^1 dx 
 \, e^{-i x (pz) } \,  {\cal P} (x, -z^2)  \   .
  \label{MPD}
\end{align}   
  Note that  Eq. (\ref{MPD})   gives a covariant definition of $x$.  
   There is no need to assume that 
$p^2=0$ or $z^2=0$ to define $x$. 
 
 \subsection{Collinear   PDFs}
 
Choosing 
some special cases  of $p$ and $z$, one can  get expressions for various parton distributions,
all in terms of the same function  ${\cal M} (-(pz), -z^2)$. 
In particular, taking  a light-like  $z$, e.g., that  
having  the light-front minus component $z_-$  only, we   
parameterize the matrix element  by  the twist-2 parton distribution $f(x)$ 
  \begin{align}
  {\cal M} (-p_+ z_- , 0)  =
   \int_{-1}^1 dx \, f(x) \, 
e^{-ixp_+ z_-} \,  \   ,  
 \label{twist2par0}
\end{align}
with $f(x)$ having  the usual interpretation of  probability that 
the parton carries  the fraction $x$ of the target momentum 
component $p_+$.  
The inverse relation is given by 
\begin{align} 
 f  (x)  =\frac{1}{2 \pi}  \int_{-\infty}^{\infty}  d\nu \, e^{-i x \nu}  \, {\cal M}(\nu , 0)    = {\cal P} (x, 0) 
  \  .
\label{fxMnu}
\end{align}  
Since $f(x) =   {\cal P} (x, 0) $, the function $ {\cal P} (x, -z^2) $ generalizes 
PDFs onto non-lightlike intervals $z^2$, and we will call it 
 {\it pseudo-PDF}.  The variable $(pz)$ is called 
 the {\it Ioffe time} \cite{Ioffe:1969kf}, 
and   ${\cal M}(\nu , -z^2) $  is    the {\it Ioffe-time distribution} \cite{Braun:1994jq}. 

Note that  the definition of $ {\cal P} (x, -z^2) $ is simpler than  that of $f(x)$ because it 
 does not require taking  a subtle $z^2 \to 0$  limit.  
  In     renormalizable theories,    
  the function  $ {\cal M} (\nu, z^2)  $ 
 has  $\sim \ln z^2$  singularities 
generating  perturbative evolution of parton densities.  
Within the   operator product expansion
(OPE)  approach,  the $\ln z^2$  singularities
are subtracted using some prescription, say, dimensional renormalization,
and the resulting PDFs   depend on the renormalization scale $\mu$, i.e., 
$f(x) \to   f(x, \mu^2)$.

 \subsection{Transverse momentum dependent
distributions}

When $z^2$ is spacelike, one can treat  $-z^2$ 
as the magnitude squared of a two-dimensional  
 vector    $ \{z_1,z_2\}$,     and introduce a two-dimensional 
 Fourier transform 
with respect to its components, i.e.,  to write 
     \begin{align}
 {\cal P} (x,  z_1^2+z_2^2) 
&   = 
    \int_{-\infty}^\infty  {d k_1 } e^{i k_1 z_1} 
   \nn & \times  
    \int_{-\infty}^\infty  dk_2 \,e^{ik_2z_2} 
      {\cal F} (x, k_1^2+k_2^2)
 \   .
  \label{PTMD}
\end{align}  
Due to rotational invariance of ${\cal P} (x,  z_1^2+z_2^2) $
in $ \{z_1,z_2\}$  plane,
the function      ${\cal F} (x, k_1^2+k_2^2)$ depends 
on ${k_1,k_2}$  through $k_1^2+k_2^2$, the fact already reflected in the notation. 
Combining this representation with Eq. (\ref{MPD}), one has 
   \begin{align}
 {\cal M} (\nu,  z_1^2+z_2^2) 
&   =  \int_{-1}^1 dx 
 \, e^{i x \nu  } \, 
    \int_{-\infty}^\infty  {d k_1 } e^{i k_1 z_1} 
   \nn & \times  
    \int_{-\infty}^\infty  dk_2 \,e^{ik_2z_2} 
      {\cal F} (x, k_1^2+k_2^2)
 \   .
  \label{MTMD}
\end{align}  

A physical interpretation of  ${\cal F} (x, k_1^2+k_2^2)$
may be given in the frame  where  the target momentum $p$ is   longitudinal,
\mbox{$p= (E, {\bf 0}_\perp, P)$,}  while  the vector  
$\{z_1,z_2\} $  is in the transverse plane.   Taking $z$ that has 
 $z_-$ and $z_\perp$  components only,  one can identify   ${\cal F} (x, k_\perp^2)$
with  the  
{\it TMD }    and write 
    \begin{align}
 {\cal P} (x,  z_\perp^2) 
&   = 
    \int  d^2{\bf k}_\perp   e^{i ({\bf k}_\perp {\bf z}_\perp)} 
      {\cal F} (x, {k}_\perp^2)
 \   .
  \label{MTMD0}
\end{align}  
In this case, the pseudo-PDFs  $ {\cal P} (x,  z_\perp^2) $ coincide
with the  {\it impact parameter distributions}, a well-known   concept 
 actively
used in TMD studies.

The   \mbox{$\sim \ln z_\perp^2$}  terms 
in   $ {\cal M} (\nu,  z_\perp^2) $ are 
produced by  the  $\sim 1/k_\perp^2$ hard tail of ${\cal  F} (x, k_\perp^2)$.
Thus,  it makes sense to visualize 
$ {\cal M} (\nu, z_\perp^2)  $   as a sum of a soft part $ {\cal M}^{\rm soft} (\nu, z_\perp^2)$, 
that has a finite $z_\perp^2\to 0$ limit  
and a  hard part     reflecting   the evolution. 
For   TMDs,    soft part  decreases faster  than $1/k_\perp^2$, say, like 
a Gaussian $e^{-k_\perp^2/\Lambda^2}$.   In  the  $z_\perp$  space, the distributions
are then concentrated in  \mbox{$z_\perp \sim 1/\Lambda$}  region. 

\section{Quasi-Distributions}

\subsection{Definition and relation to TMDs}

Since one cannot have light-like separations on the lattice,
it was proposed  \cite{Ji:2013dva}  to consider 
spacelike separations  
$z= (0,0,0,z_3)$ [or, for brevity, \mbox{$z=z_3$}]. Then, 
 in  the   \mbox{$p=(E, 0_\perp, P)$}  frame,
 one  introduces the quasi-PDF  $Q(y, P)$    through a parametrization  
  \begin{align}
  \langle p |   \phi(0) \phi (z_3)|p \rangle 
=  & 
\int_{-\infty}^{\infty}   dy \, 
 Q(y, P) \,  e^{i y  P z_3 } \, 
 \  . 
 \label{newVDFxzQ}
\end{align} 
According to  this definition, the function $Q(y,p)$ characterizes  the probability 
that  the  parton  carries fraction $y$ of hadron's third momentum
component $P$.  Viewing  the matrix element as a function 
of the $\nu$ and $-z^2$ variables
(they are given by $Pz_3$ and $z_3^2$ in this case), we have 
 \begin{align}
{\cal M} (\nu,z_3^2) 
=  & 
\int_{-\infty}^{\infty}   dy \, 
 Q(y, P) \,  e^{i y  \nu } \, 
 \  . 
 \label{newVDFxzQ}
\end{align} 
Noticing that $z_3^2= \nu^2/P^2$, we get  the inverse  Fourier transformation 
in the form 
\begin{align} 
  Q(y,  P)   =\frac{1}{2 \pi}  \int_{-\infty}^{\infty}  d\nu \, 
   \, e^{-i y  \nu}  \, {\cal M} (\nu,  \nu^2/P^2)    \  .
\label{IxM}
\end{align}  
It 
indicates  that $  Q(y,  P)  $  tends to $f(y)$  in the 
$P \to \infty$  limit,  as far as    ${\cal M} (\nu,  \nu^2/P^2) \to {\cal M} (\nu, 0)$.

Thus, the deviation of the quasi-PDF $Q(y,  P)$ from  the PDF $f(y)$
is determined by the dependence of ${\cal M} (\nu,  z_3^2)$
with respect to its second argument.
By  \mbox{Eq. (\ref{MTMD}),}  this dependence is related to the dependence
of the TMD  $ {\cal F} (x, \kappa^2)$ on its second argument $\kappa^2$.
Hence,  the difference between  $Q(y,  P)$ and  $f(y)$
may be   described in terms of TMDs. 

To this end, we  
incorporate the fact that   \mbox{Eq. (\ref{MTMD}) } 
is a {\it mathematical}  relation between 
  the function  \mbox{${\cal M} (\nu,  z_1^2+z_2^2) $} 
and  the function      $ {\cal F} (x, k_1^2+k_2^2)$, 
no matter what is a  {\it physical}  meaning of the variables 
$z_1,z_2$ and $k_1,k_2$. 
Thus, we
 substitute  \mbox{Eq. (\ref{MTMD})}   with  $z_1=0$ and $z_2 =\nu/P$ 
into  Eq. (\ref{IxM})  to  convert it   into the  
expression for  quasi-PDFs in terms of TMDs 
  \begin{align}
 Q(y, P) /P =  & \,\int_{-\infty}^{\infty} d  k_1
\int_{-1} ^ {1} d  x \,   {\cal F} (x, k_1^2+(y-x)^2P^2 )
  \  . 
 \label{QTMD}
 \end{align} 
 
 Originally, this relation was derived in Ref.  \cite{Radyushkin:2016hsy} using
 a Nakanishi-type representation of Refs.   \cite{Radyushkin:2014vla,Radyushkin:2015gpa}.
 Now, we see that it is a mere consequence of  Lorentz invariance.

\subsection{Quantum chromodynamics (QCD)  case}  

The formulas derived above are
directly   applicable for non-singlet parton densities   in QCD.
In that case,  one    deals with matrix elements  of  
    \begin{align}
 {\cal M}^\alpha  (z,p) \equiv \langle  p |  \bar \psi (0) \,
 \gamma^\alpha \,  { \hat E} (0,z; A) \psi (z) | p \rangle \  
\label{Malpha}
\end{align}
type, where  $
{ \hat E}(0,z; A)$ is  the  standard  $0\to z$ straight-line gauge link 
 in the quark (fundamental) representation.
 These matrix elements   may be decomposed into $p^\alpha$ and $z^\alpha$ parts:
\begin{align} 
{\cal M}^\alpha  (z,p) = &2 p^\alpha  {\cal M}_p (-(zp), -z^2) 
\nn & + z^\alpha  {\cal M}_z (-(zp),-z^2)
\ .
\end{align}
The ${\cal M}_p (-(zp), -z^2) $ part gives the twist-2 distribution when  $z^2 \to 0$,
while $ {\cal M}_z ((zp),-z^2)$ is a  purely higher-twist contamination,
and it  is better to  get rid of it. 

If one takes  $z=(z_-, z_\perp)$ in the $\alpha=+$  component of 
${\cal M}^\alpha$,  the  $z^\alpha$-part drops out,  
and one can introduce a TMD  ${\cal F}(x, k_\perp^2)$  that 
is  related to  $ {\cal M}_p (\nu, z_\perp^2)$  by the scalar formula 
(\ref{MTMD}). 
For  quasi-distributions,  
the easiest way to remove   the $z^\alpha$ contamination is to take the
  time component of \mbox{$ {\cal M}^\alpha  (z=z_3,p)$}  and define
 \begin{align}
&  {\cal M}^0   (z_3,p)   
  =  2p^0 
 \int_{-1}^1 dy\,  
Q(y,P) \, 
 \,  e^{i  y Pz_3 }  \  . 
 \label{OPhixspin12}
\end{align} 
Then the  connection between $Q(y,P)$   and 
 ${\cal F}(x, k_\perp^2)$ is given 
by the scalar  formula (\ref{QTMD}).

One may notice that  the operator  defining $ {\cal M}^\alpha  (z,p) $
involves a straight-line link from $0$ to $z$ rather than a
stapled link usually used in the definitions of TMDs appearing  in the description of
Drell-Yan and semi-inclusive DIS processes.   As is well-known, the stapled links 
reflect initial or final state interactions inherent in  these processes.
The ``straight-link'' TMDs, in this sense, describe the structure
of a hadron when   it   is in 
 its  non-disturbed or ``primordial''  state. While it is unlikely that such a TMD can be measured
 in a scattering experiment, it is a well-defined  QFT object,
 and one may hope that it can be measured on the lattice.

 \subsection{Momentum distributions}

  The quasi-PDFs describe the distribution in   the fraction $y \equiv k_3/P$ of the 
 third component  $k_3$ of the  parton momentum to that of the hadron. 
 One can  introduce distributions  in $k_3$ itself:
 $R (k_3,P) \equiv Q(k_3/P,P)/P$.
 Then we can rewrite Eq. (\ref{QTMD}) as
  \begin{align}
 R(k_3, P)  =  & \,\int_{-\infty}^{\infty} d  k_1
\int_{-1} ^ {1} d  x \,   {\cal F} (x, k_1^2+(k_3-xP)^2 )
  \  
 \label{RTMD0}
 \end{align} 
 or, switching to   the linear argument \mbox{$k_3-xP$,} 
   \begin{align}
 R(k_3, P)  =  &  \int_{-1}^1 dx\,   {\cal R} (x, k_3-xP)
  \  , 
 \label{RTMD}
 \end{align} 
 where
  \begin{align}
 {\cal R} (x, k_3  )   \equiv   & \,\int_{-\infty}^{\infty} d  k_1
  {\cal F} (x, k_1^2+k_3^2)
  \  
 \label{calD}
\end{align} 
is the TMD $ {\cal F} (x, \kappa^2)$  integrated over the $k_1$   component
of 
the  two-dimensional  vector $\kappa=\{k_1,k_3\}$. 
According to (\ref{calD}),  ${\cal R} (x, k_3  ) $ depends on $k_3$ through $k_3^2$. 

For a hadron at rest, we have
  \begin{align}
 R(k_3, P=0)  \equiv   &\, r(k_3)=
   \int_{-1}^1 dx\,   {\cal R} (x, k_3  )
  \  .
 \label{DTMDsmall}
\end{align}

This one-dimensional distribution may be  directly  obtained  through   a  parameterization  
of  the density 
 \begin{align}
 \rho (z_3^2) \equiv & {\cal M} (0,z_3^2) =
\int_{-\infty}^{\infty}   dk_3 \, 
 r(k_3)  \,  e^{i k_3 z_3 } \,  
 \
 \label{dk3}
\end{align} 
given by    $  \langle p |   \phi(0) \phi ( z_3 )|p \rangle |_{ {\bf p} = {\bf 0} } $. 
Thus, $r(k_3)$  describes a primordial distribution of $k_3$ 
(or any other  component of  ${\bf k}$)  in  
a  rest-frame hadron.


 The  formula (\ref{RTMD}) has a  straightforward  interpretation. 
According to it,  when the hadron is moving, the parton's $k_3$ momentum
 has  two sources.
 
 The first part, $xP$ comes from the motion
of the hadron as a whole, and   the probability to get $xP$ is governed by the dependence of
 the TMD  ${\cal F} (x, \kappa^2)$ on its first argument, $x$.
 
 On the other hand, the  probability to get the remaining part $k_3-xP$   
is governed by the  dependence of the TMD  on its second argument,
$\kappa^2$, 
describing    the primordial 
 rest-frame 
 momentum distribution.   
 
 The parameter $x$ appears in both arguments of 
 $ {\cal R} (x, k_3-xP)$  in Eq. (\ref{RTMD}), 
 i.e., $R(k,P)$   is given by a convolution. 
  In this sense, the momentum distributions  $R(k,P)$ 
 and, hence, the quasi-PDFs  have a hybrid structure  influenced by 
 the shape  both of 
 PDFs  and  rest-frame distributions.

\subsection{Factorized models.} 
 
 Since  the two sources of $k_3$  look like   independent, it is natural
to demonstrate  the hybrid nature of momentum distributions and quasi-PDFs using 
 a factorized model \mbox{$ {\cal R} (x, k_3-xP) = f(x) r (k_3-xP)$} 
(the $x$ integral of $f(x)$ is  normalized \mbox{to 1).}
For  original ${\cal M} (\nu,-z^2)$ function,  this Ansatz corresponds 
to the factorization assumption ${\cal M} (\nu,-z^2) = {\cal M} (\nu,0){\cal M} (0,-z^2)$.

\begin{figure}[t]
    \centerline{\includegraphics[width=3in]{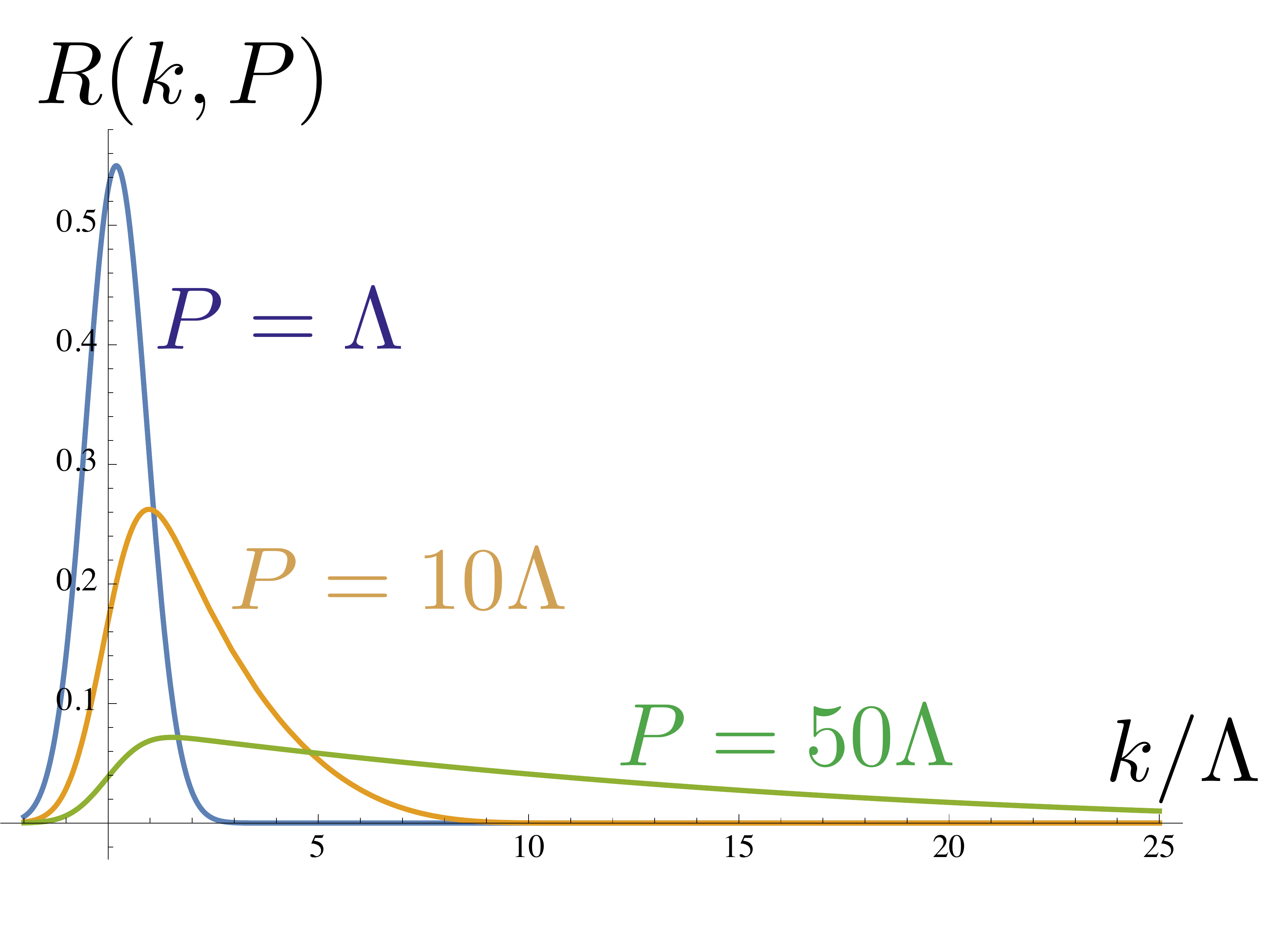} }
    \vspace{-5mm} 
    \caption{Momentum distributions  $R(k,P)$  in  the factorized Gaussian model   for {$P/\Lambda =1,10,50$}.}
    \label{R}
    \end{figure}

For illustration,    we take   a Gaussian  form  
$
\rho_G(z_3^2)=  e^{-z_3^2\Lambda^2/4} 
$ for the rest-frame density.  It corresponds to 
  \begin{align}
r_G(k_3) =  & 
\frac{1}{\sqrt{\pi } \Lambda}  e^{-k_3^2/\Lambda^2}
  \  .
 \label{DTMDsmallfac}
\end{align} 
For $f(x)$, we take 
 a simple PDF resembling  nucleon valence densities
$f(x)=4(1-x)^3 \theta (0\leq x \leq 1)$.     
As one can see from Fig. \ref{R}, the curve for $R(k,P)$ 
changes from a Gaussian shape for small $P$  to a shape resembling 
 stretched PDF  for large $P$.  
 
This result  is in perfect compliance 
with a known fact   that wave functions 
of moving hadrons are not given by a mere kinematical ``boost''
of the rest-frame wave functions. 
Indeed, with increasing  $P$,  the  impact  of  the
rest-frame distribution $r(k)$  
 is  less and less visible, and  eventually the shape of 
 $R(k,P)$  is determined by a completely 
 different function $f(k/P)$.

\begin{figure}[t]
    \centerline{\includegraphics[width=3.1in]{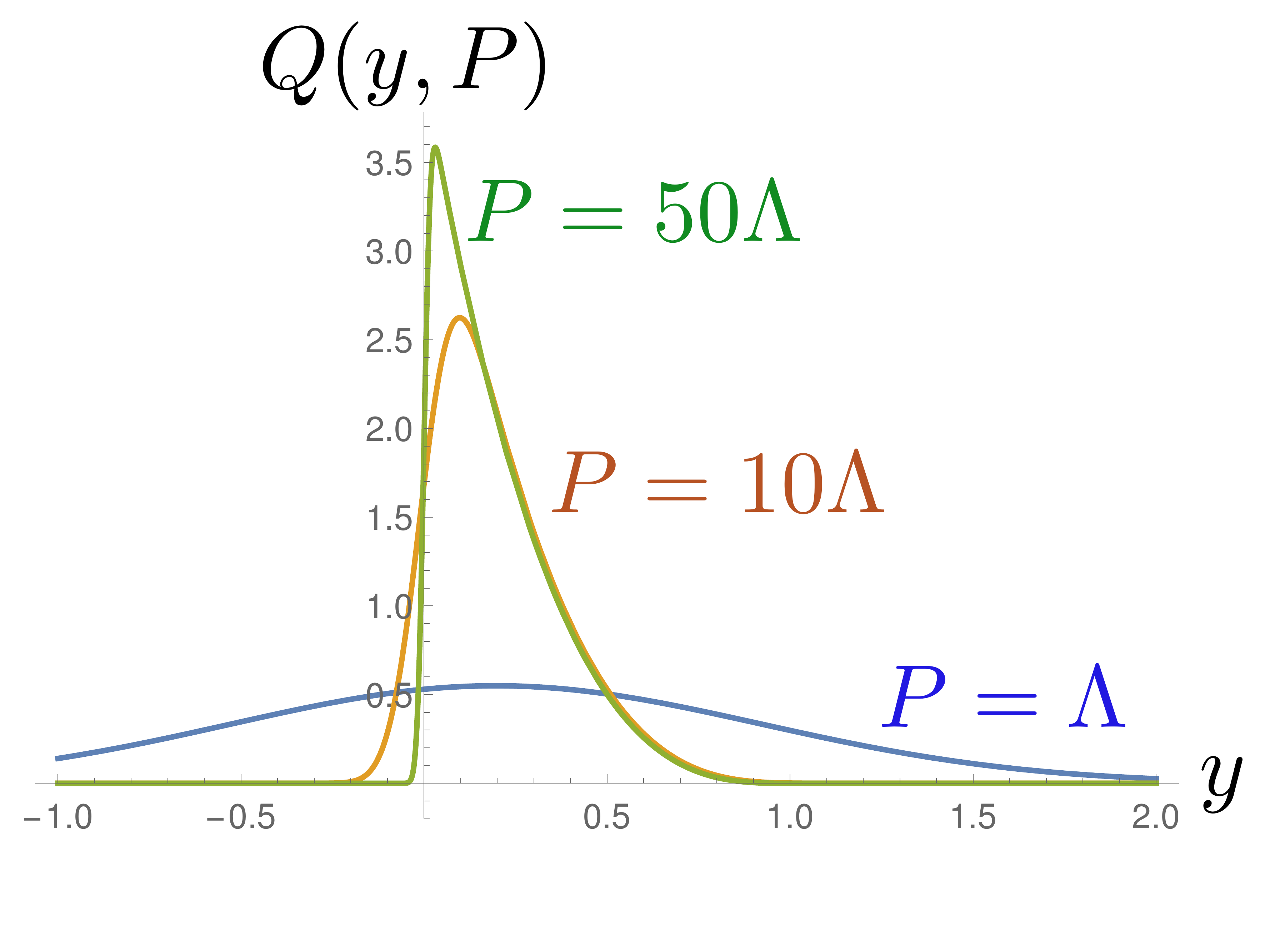}}
    \vspace{-0.7cm}
    \caption{Evolution of quasi-PDF $Q(y,P)$  in the  factorized Gaussian  model  for {$P/\Lambda =1,10,50$}.
    \label{Qgy}}
    \end{figure}

 Rescaling to the $y=k/P$ variable 
 gives the quasi-PDF $Q(y,P)$ shown in Fig. \ref{Qgy}. 
 For large $P$, it clearly tends to the $f(y)$   PDF  form.
In particular,   using a momentum 
$P \sim 10 \Lambda$  one gets  a quasi-PDF that is rather close to the $P \to \infty$ limiting shape. 
Still, since  $\Lambda \sim \langle k_\perp \rangle$,    assuming  the folklore value 
$ \langle k_\perp \rangle \sim $ 300 MeV  one   translates  the 
 \mbox{$P\sim 10 \Lambda$}  estimate into $P \sim 3$ GeV,
which is  uncomfortably  large. Thus, a natural question is  how to   improve the convergence.

\subsection{Pseudo-PDFs}  

A  formal  reason for the   complicated structure  of a quasi-PDF
$Q(y,P)$ 
is the fact that it is  obtained by the  $\nu$-integral of 
${\cal M} (\nu, z_3^2)e^{i \nu y}$ 
 along a non-horizontal line $z_3=\nu/P$ 
in  the $(\nu, z_3)$  plane (see Eq. (\ref{IxM})).  With increasing $P$, its slope 
decreases,  the line becomes more  horizontal,
and quasi-PDFs convert into PDFs.

In contrast,  pseudo-PDFs ${\cal P} (x,z_3^2)$,
by definition,  are given by integration of ${\cal M} (\nu, z_3^2)e^{i \nu x}$  over 
horizontal lines $z_3=$ const.  A very attractive feature of the  pseudo-PDFs is that they 
  have 
the $-1\leq x \leq 1$ support  for all $z_3$ values.  For small $z_3$, 
they   convert  into PDFs.

 More precisely,  when $z_3$  is small, $1/z_3$
 is analogous to the renormalization parameter $\mu$ 
of scale-dependent PDFs $f(x,\mu^2)$ of the standard OPE approach.  

There is a  subtlety, however, that  while the \mbox{$\mu^2$-dependence } 
of PDFs $f(x,\mu^2)$ comes  solely from the evolution logarithms  $\ln (\mu^2/m^2)$,
the $z_3^2$-dependence of quasi-PDFs in QCD comes both from
the evolution logarithms  $\ln (z_3^2 m^2)$
and from the ultraviolet logarithms  $\ln (z_3^2 \mu_R^2)$,
where $\mu_R$ is a   cut-off parameter for divergences related to the gauge link
renormalization (see Ref. \cite{Craigie:1980qs}). 
At the leading logarithm level, these divergences do not depend on $\nu$.
As a result, the ``reduced''  Ioffe-time distribution 
 \begin{align}
{\mathfrak M} (\nu, z_3^2) \equiv \frac{ {\cal M} (\nu, z_3^2)}{{\cal M} (0, z_3^2)} \  
 \label{redm}
\end{align}
satisfies,  
 for small $z_3$,  the   leading-order evolution equation   
     \begin{align}
    \frac{d}{d \ln z_3^2} \,  
{\mathfrak M} (\nu, z_3^2)    = - \frac{\alpha_s}{2\pi} \, C_F 
\int_0^1  du \,   B ( u )  \,   {\mathfrak  M} (u \nu, z_3^2)  
\label{74}
 \end{align}
 with respect to $1/z_3$ 
  that 
is similar to  the evolution equation for $f (x,\mu^2)$ with respect to $\mu$. 
The leading-order evolution kernel $B(u)$ for the non-singlet quark case
is given  \cite{Braun:1994jq} by
   \begin{align}
B (u)    &=  \left [ \frac{1+u^2}{1- u} \right ]_+  \ 
 , 
\label{74a}
 \end{align} 
 with $[ \ldots ]_+$  denoting the standard ``plus'' prescription.

\begin{figure}[t]
    \centerline{\includegraphics[width=3.3in]{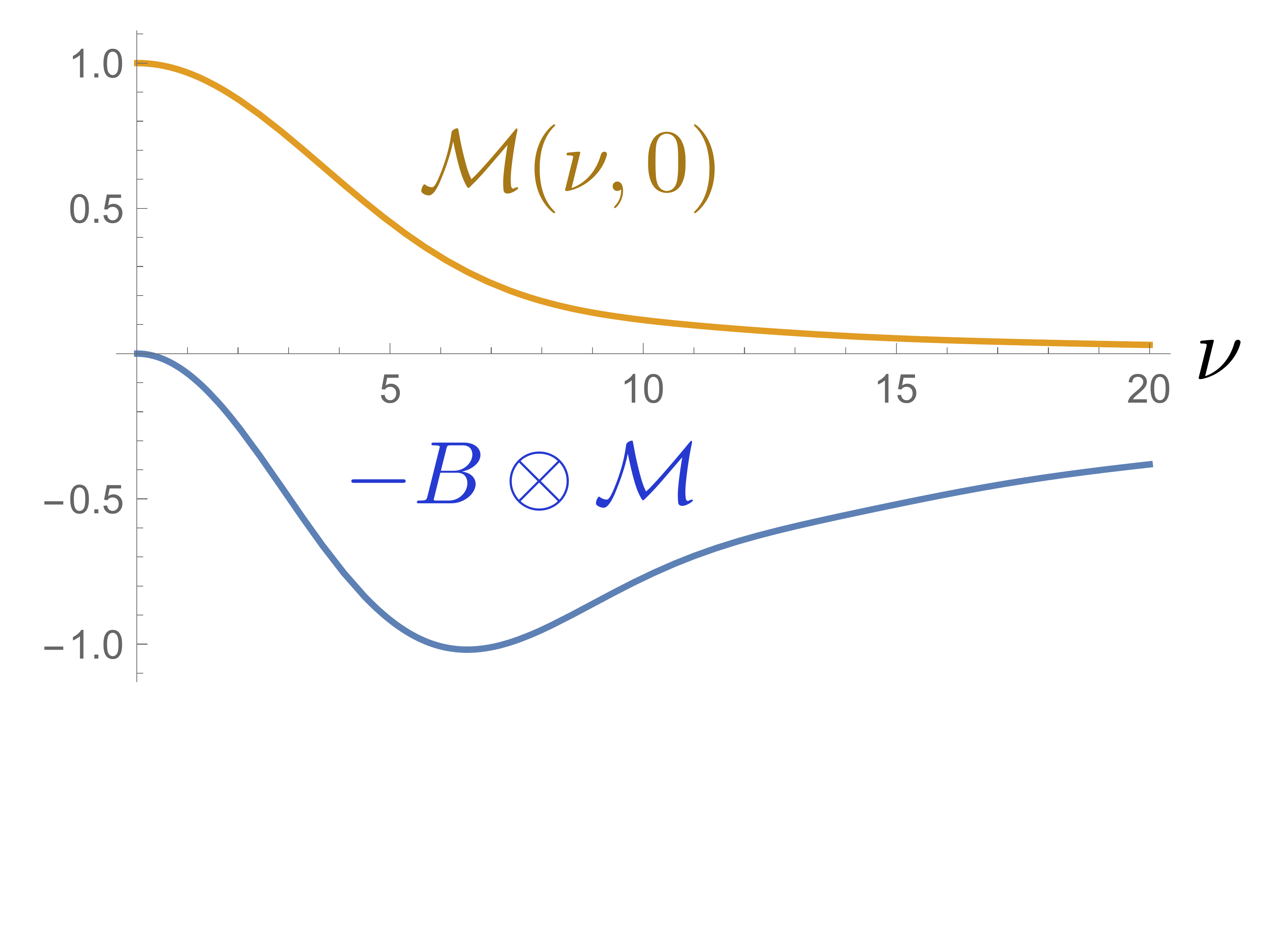}}
    \vspace{-0.6cm}
    \caption{Model Ioffe-time distribution ${\cal M} (\nu,0)$  and
    the function $B \otimes {\cal M}$  governing its evolution.
    \label{MBM}}
    \end{figure}

 For  the model used above (and $x \to -x$ symmetrized, as required for
 non-singlet PDFs), we have
 ${\cal M} (\nu, 0) ={12}\left [\nu^2 - 4\sin ^2( \nu/2) \right ]  /{\nu^4}    \  .
$
The shape of this function and of the convolution integral
$B \otimes {\cal M}(\nu)$   are  shown in \mbox{Fig. \ref{MBM}. } 
As one can see, $B \otimes {\cal M}(\nu)$  vanishes for 
$\nu=0$, which reflects  conservation of the vector current. 
Thus, the rest-frame density ${\mathfrak M} (0,z_3^2)$  is not affected 
by perturbative evolution.

\subsection{Lattice implementation} 

A possible way to find the Ioffe-time distributions  on the lattice
(suggested by K. Orginos)   is to 
calculate ${\cal M} (Pz_3, z_3^2)$  for several values of $P$,
and then to fit the results  by  a function of $\nu$ and $z_3^2$. 
 
Recalling  our discussion of two apparently independent
sources of obtaining $k_3$  for a moving hadron, one may hope 
that 
${\cal M} (\nu, z_3^2)$ factorizes, i.e.,  
 ${\cal M} (\nu, z_3^2)= {\cal M} (\nu, 0) {\cal M} (0, z_3^2)$.    Then 
 the reduced function  ${\mathfrak M} (\nu, z_3^2) $ defined by Eq. (\ref{redm}) 
is equal to $ {\cal M} (\nu, 0)$, 
and   
the  goal of obtaining ${\cal M} (\nu, 0)$ is reached.
Formally, what remains is just to take its Fourier transform to get the PDF $f(x)$.

In fact, such a factorization has been already observed 
several years ago 
in the pioneering study \cite{Musch:2010ka} of the 
transverse momentum distributions in  lattice QCD.

A serious  disadvantage of quasi-PDFs is that they have the $x$-convolution
structure (\ref{QTMD})  even   in  a  favorable situation when 
the TMD  [and ${\cal M} (\nu, z_3^2)$]    factorizes.
On the other hand, using pseudo-PDFs in the form of
 the ratio ${\mathfrak M} (\nu, z_3^2)$,  one 
  divides  out the $z_3^2$-dependence of the
 primordial distribution  without affecting the $\nu$-dependence
 that dictates the shape of PDF.  
 
 A further advantage  of using the ratio (pointed out by K. Orginos)  is 
the   cancellation of the   \mbox{$z_3$-dependence}   
 generated 
by the lattice  renormalization of 
the gauge link  ${ \hat E} (0,z_3; A)$. 
Such a renormalization is required by 
linear $|z_3| \delta m$ (where $\delta m \sim 1/a$, and $a$ is the ultraviolet cut-off)
and logarithmic $\ln (z_3^2/a^2)$ divergences  \cite{Polyakov:1980ca,Dotsenko:1979wb}. 
Due to their local nature, they  are expected to 
combine into a $\nu$-independent factor
$Z(z_3^2/a^2)$ that is   the same in the numerator and denominator 
of the ratio  ${\mathfrak M} (\nu, z_3^2) $.

The   multiplicative renormalizability  of the linear 
divergences of ${\cal M} (\nu, z_3^2)$
to all orders 
was recently argued  in Refs. \cite{Ishikawa:2016znu,Chen:2016fxx}. 
A general proof for both linear and logarithmic divergences was claimed  in 
Ref. \cite{Ishikawa:2017faj} 
on the basis of  a direct analysis of relevant  Feynman graphs.

Another approach  \cite{Ji:2017oey,Green:2017xeu}  is to treat 
$ \hat E(0,z;A) $ as 
 $ h(0) \bar h (z) $,
 where the auxiliary field $h(z)$  is  analogous to the infinitely heavy
 quark field of the heavy quark effective  theory (HQET).
 Since HQET is known to be multiplicatively renormalizable \cite{Bagan:1993zv}
 this means that $\bar \psi(0)  \hat E(0,z;A) \psi (z)$ is also 
 multiplicatively renormalizable to all  orders in perturbation theory.

In reality,  ${\mathfrak M} (\nu, z_3^2)$   will have 
a  residual \mbox{$z_3^2$-dependence.} 
 It comes   both 
from  a  possible 
violation of factorization for  the soft part  
(according to results of Ref. \cite{Musch:2010ka}, 
it is expected to be rather mild) and 
from  mandatory  perturbative  evolution. For a  nonzero $\nu$, 
the latter should be visible as a
$\ln (1/z_3^2 \Lambda^2)$  spike for small $z_3^2$.

Hence,  a proposed   strategy is  to extrapolate 
${\mathfrak M} (\nu, z_3^2)$ to $z_3^2=0$ from not too small
values of $z_3^2$, say,  from those above 0.5 fm$^2$.
 The resulting function  ${\cal M} ^{\rm soft}(\nu,0)$
may be treated as 
the Ioffe-time distribution producing the PDF $f_0(x)$ 
 ``at low normalization point''.  The remaining 
$\ln (1/z_3^2 \Lambda^2)$  spikes  at small $z_3$ will generate its evolution.

To  convert  ${\mathfrak M} (\nu, z_3^2)$ into a function of $x$, one should, in principle,
know  ${\mathfrak M} (\nu, z_3^2)$ for all $\nu$, which is  impossible.
The maximal values  of $\nu$ reached in existing lattice calculations 
range from $3\pi $ \cite{Lin:2014zya}     to $5\pi$   \cite{Alexandrou:2015rja}   and $6 \pi $  \cite{Orginos:2017kos}. 
Taking a Fourier transform in these limited ranges produces 
unphysical oscillations in $x$. 
Thus, the  idea is to avoid  the Fourier transform in $\nu$, 
and  just compare the reduced  Ioffe-time distributions obtained from the
 lattice   
with those derived from experimentally known parton distributions.

Of course, an actual technical  implementation of this program 
should be discussed when the lattice data on ${\mathfrak M} (\nu, z_3^2)$
will become  available.

\section{Summary}

In this paper, we  showed that  quasi-PDFs may be  seen  as hybrids of PDFs and the 
primordial rest-frame 
momentum distributions of partons. 
In this context,  the parton's $k_3$ momentum
comes from the motion
of the hadron as a whole and from  the primordial 
 rest-frame 
 momentum distribution.    The  complicated convolution  nature of quasi-PDFs 
necessitates  using  $p_3 \gtrsim $ \mbox{ 3 GeV}  to wipe out
the primordial momentum distribution effects and 
 get reasonably close to the PDF limit.
 
As an alternative approach, we propose   to use 
pseudo-PDFs  $ {\cal P} (x, z_3^2)  $ that generalize
the light-front PDFs   onto spacelike intervals.   By a  Fourier transform,
they are related to the 
 Ioffe-time  distributions  ${\cal M} (\nu, z_3^2) $  given by  generic matrix elements  
 written as  functions of $\nu = p_3 z_3$ and $z_3^2$. 
 The advantageous features  of  pseudo-PDFs  are  that they, first,  have the 
 same $-1 \leq x  \leq 1$ support  as PDFs, and second, their 
 $z_3^2$-dependence  for small $z_3^2$  is governed by  a usual  evolution equation.
 
 Forming   the ratio ${\cal M} (\nu, z_3^2)/{\cal M} (0, z_3^2)$  
 of  Ioffe-time  distributions  one 
 divides   out the bulk of $z_3^2$   dependence generated by  
 the primordial rest-frame distribution. 
 Furthermore,  taking this ratio 
 one can exclude the \mbox{ $z_3^2$-dependent}   factor coming from the 
 lattice renormalization of the $\hat E (0,z_3;A)$ link 
creating    difficulties (see, e.g.,  \cite{Ishikawa:2016znu}) for   lattice calculations of quasi-PDFs. 

Testing the efficiency of using pseudo-PDFs for  lattice extractions 
of PDFs is a  challenge for future studies.  

In fact, while this paper was in the review
process, an actual lattice calculation  \cite{Orginos:2017kos} based on the ideas of  
 the present  paper was performed.
 It  has clearly demonstrated  the presence of a linear  component in the $z_3$-dependence 
    of  the rest-frame function ${\cal M} (0,z_3^2)$, that may be attributed 
    to the $Z(z_3^2/a^2) \sim e^{-c|z_3|/a}$ behavior 
    generated by the gauge link. It was also observed that 
      the ratio ${\cal M} (Pz_3 , z_3^2)/{\cal M} (0, z_3^2)$
    has a Gaussian-type  behavior with respect to $z_3$, which indicates that 
    the $Z(z_3^2/a^2)$ factors entering into the numerator and denominator
    of the ${\mathfrak  M} (Pz_3, z_3^2)$  ratio have been canceled, as we expected.

 Furthermore, it was  found  that when plotted as a function of $\nu$ and $z_3$,
  the data for the reduced distribution ${\mathfrak  M} (\nu, z_3^2)$  have a very mild dependence 
  on $z_3^2$.   This observation indicates  that 
    the soft part of the \mbox{$z_3^2$-dependence}  of $   {\cal M} (\nu, z_3^2)$ 
    has been  canceled by the rest-frame density $   {\cal M} (0, z_3^2)$.
    This phenomenon corresponds to factorization of the $x$- and $k_\perp$-dependence
    for the  soft part of the TMD ${\cal F} (x, k_\perp^2)$.

     It was also demonstrated that the residual  \mbox{$z_3$-dependence} 
     for small $z_3 \leq 4 \,a$, 
    may be explained by  perturbative evolution, with the $\alpha_s$ value
     corresponding to  $\alpha_s/\pi =0.1$.

\acknowledgements


I thank C.E. Carlson for  his 
 interest in  this work, and also V. M.  Braun and X. Ji for discussions
 and suggestions.
  I am especially
grateful to K. Orginos for stimulating discussions and suggestions concerning 
 the lattice implementation of the  approach. 
This work is supported by Jefferson Science Associates,
 LLC under  U.S. DOE Contract \#DE-AC05-06OR23177
 and by U.S. DOE Grant  \mbox{\#DE-FG02-97ER41028. }

\end{document}